# Fast-neutron irradiation response of β-$Ga_2O_3$ Schottky barrier diodes on sapphire substrate

Saleh Ahmed Khan[1], Ahmed Ibreljic[1], Sourav Sarker[1], Stephen Margiotta[1], and Anhar Bhuiyan[1, a)]

[1]Department of Electrical and Computer Engineering, University of Massachusetts Lowell, MA 01854, USA

[a)] Corresponding author Email: anhar_bhuiyan@uml.edu

## Abstract

In this work, we investigate the electrical response of Ni/β-$Ga_2O_3$ Schottky barrier diodes (SBDs) fabricated on c-plane sapphire substrates under fast-neutron irradiation at a fluence of $1\times10^{15}$ n·cm$^{-2}$. The β-$Ga_2O_3$ heteroepitaxial device stack is grown by low-pressure chemical vapor deposition (LPCVD) and comprises a 1.05 μm unintentionally doped buffer, a 2.10 μm n$^+$ layer ($N_{D,\mathrm{n+}} \approx 1 \times 10^{19}$ cm$^{-3}$), and a 3.15 μm drift layer ($N_{D,\mathrm{drift}} \approx 2.07 \times 10^{17}$ cm$^{-3}$), with plasma-free mesa definition achieved using Ga-assisted LPCVD-based etching. Prior to irradiation, the devices exhibit a turn-on voltage of 1.20 V, minimum differential specific on-resistance $R_{on,sp}$ = 8.43 mΩ·cm², ideality factor η = 1.32, and a Schottky barrier height of 1.29 eV. Even after exposure to a high 1 MeV equivalent fast-neutron fluence of $1\times10^{15}$ n·cm$^{-2}$, the devices remain operational, although the forward current decreases, the turn-on voltage increases to 2.40 V, and minimum $R_{on,sp}$ rises to 8.85 mΩ·cm². Simultaneously, the barrier height increases to 1.34 eV and the net donor concentration in the drift layer decreases by ~ 46 percent to $1.12\times10^{17}$ cm$^{-3}$, corresponding to a carrier-removal rate of ~ 95 cm$^{-1}$. Temperature-dependent measurements from 25 to 250 °C confirm that thermionic emission remains the dominant transport mechanism and show a systematic suppression of reverse leakage. The breakdown voltage increases from 101 V to 135 V, accompanied by a reduction in the estimated peak electric field from 2.75 to 2.34 MV·cm$^{-1}$, consistent with neutron-induced drift-layer compensation. Silvaco Two-dimensional technology computer-aided design (TCAD) simulations reveal that the electric field becomes more uniformly distributed over a wider depletion region, reducing field crowding at the Schottky edge. The measured behavior offers critical insight into neutron-induced donor compensation in β-$Ga_2O_3$ films grown on non-native substrates and demonstrates the potential of heteroepitaxial β-$Ga_2O_3$ SBDs to maintain stable operation under high-fluence neutron environments relevant to space and nuclear electronics.

## I. INTRODUCTION

β-$Ga_2O_3$ has rapidly established itself as one of the most promising ultrawide bandgap semiconductors for high-voltage and high-power electronics, owing to its large bandgap (~4.8 eV), high critical electric field (6-8 MV·cm$^{-1}$) [1, 2], and the commercial availability of large-area melt-grown substrates [3]. These attributes enable device architectures, both lateral and vertical [4-17], that can sustain multi kilovolt breakdown voltages. Although epitaxy on native substrates has driven much of the early progress through techniques such as metal-organic chemical vapor deposition (MOCVD), molecular beam epitaxy (MBE), halide vapor phase epitaxy (HVPE), and low-pressure chemical vapor deposition (LPCVD) [18-33], heteroepitaxial β-$Ga_2O_3$ on sapphire is emerging as a promising alternative direction due to its potential advantages in material cost, electrical insulation, mechanical robustness, and compatibility with wafer-scale processing [34-38]. Of particular interest is the LPCVD platform, which enables high-quality β-$Ga_2O_3$ heteroepitaxy with high growth rates [34, 37, 39, 40] as well as provides *in situ* plasma-free Ga-assisted etching [38, 41, 42] within the same reactor. This unified approach eliminates ion-induced surface damage, preserves sidewall integrity, and provides a scalable path toward vertical device topologies on non-native substrates. Such scalable β-$Ga_2O_3$ platforms are particularly attractive for emerging high-power systems intended for operation in radiation-intensive and extreme environments, where both electrical robustness and manufacturability are essential.

As interest in β-$Ga_2O_3$ power electronics expands to space systems, nuclear detection instrumentation, and radiation-intensive extreme environments, understanding the radiation response of this material system has become crucial [43, 44]. Fast neutrons, in particular, pose a significant challenge because they induce displacement cascades that generate vacancies, interstitials, and defect complexes [45, 46] capable of compensating shallow donors in β-$Ga_2O_3$ [47, 48]. Past studies on neutron-irradiated devices grown on native substrates have demonstrated high carrier-removal rates, degradation in forward characteristics, and modifications to breakdown behavior [47, 49-51]. At the same time, β-$Ga_2O_3$ has shown an excellent resilience: devices often preserve rectification, avoid catastrophic leakage [49, 50], and exhibit partial performance recovery through thermal or electro-thermal annealing [45, 46, 52]. However, despite growing interest in β-$Ga_2O_3$ for radiation-hard power technologies, prior neutron studies have focused almost exclusively on devices fabricated on native substrates. In parallel, heteroepitaxial β-$Ga_2O_3$ on sapphire has gained increasing relevance as a scalable and low-damage device platform, an approach supported by the long-standing success of GaN-on-sapphire technologies [53, 54]. To date, however, the neutron response of β-$Ga_2O_3$ devices fabricated on sapphire has not been systematically examined, particularly for structures realized using LPCVD-based heteroepitaxy and plasma-free Ga-assisted etching. This fully CVD-based processing route is of growing interest for vertical device architectures on non-native substrates, where scalable and cost-effective fabrication platforms are required for both space and terrestrial power electronics.

This work addresses this critical gap by providing a systematic investigation of fast-neutron irradiation effects in β-$Ga_2O_3$ Schottky barrier diodes grown and etched by LPCVD on c-plane sapphire substrates under a high 1 MeV equivalent neutron fluence of $1\times10^{15}$ n $cm^{-2}$. By leveraging a plasma-free growth and etching process, this device platform enables an isolated view of neutron-induced bulk compensation effects without the confounding influence of plasma damage or interface deterioration, while also providing insight into device stability under severe radiation conditions relevant to space, nuclear, and defense electronics. We examine the evolution of electrical characteristics, including forward conduction, on-resistance, Schottky barrier height, capacitance profiles, temperature-dependent transport, and reverse-blocking capability, before irradiation, immediately after exposure to a 1 MeV equivalent fluence of $1\times10^{15}$ n $cm^{-2}$, and following controlled electro-thermal annealing.

## II. EXPERIMENTAL DETAILS

The β-$Ga_2O_3$ epitaxial heterostructure was grown in a custom horizontal low-pressure chemical vapor deposition (LPCVD) reactor on commercially available c-plane sapphire substrates with a 6° miscut in order to promote step-flow growth and suppress extended defect formation [37]. Ultra-high-purity argon served as both carrier and purge gas, while high purity oxygen and solid metallic gallium pellets were used as the O and Ga precursors, respectively. The total reactor pressure during growth was maintained near 1.5 Torr and the substrate temperature was fixed at 1000 °C, with the substrate positioned 7 cm downstream from the gallium source boat. The epitaxial stack consisted of three electrically distinct layers with a total thickness of 6.3 μm as shown in. Figure 1 (a). Growth started with a 1.05 μm thick unintentionally doped β-$Ga_2O_3$ buffer layer. This was followed by a 2.10 μm thick heavily doped $n^+$ β-$Ga_2O_3$ layer grown using a $SiCl_4$ flow of 0.5 sccm to achieve a donor concentration of $1\times10^{19}$ $cm^{-3}$, which served as the common contact layer for the vertical quasi-Schottky structure. Finally, a 3.15 μm moderately doped drift layer was deposited by reducing the $SiCl_4$ flow to 0.001 sccm to achieve a target donor density of $2.07\times10^{17}$ $cm^{-3}$. After epitaxy, mesa isolation and quasi-vertical current path definition were realized using our Ga-assisted LPCVD etching process. A 100 nm thick $SiO_2$ hard mask was deposited by plasma enhanced CVD and patterned photolithographically to expose the regions to be recessed. The wafers were then transferred back into the LPCVD reactor, where thermally activated etching proceeded at 1050 °C and 1.2 Torr under oxygen deficient conditions with argon carrier gas and an upstream metallic Ga source. The etch depth was chosen to be 3.6 μm so that the entire 3.15 μm drift layer was recessed within the mesa region and a portion of the underlying $n^+$ layer was exposed between mesas. Following mesa formation and removal of the $SiO_2$ etch mask in a dilute buffered oxide etch (BOE), device fabrication proceeded through sequential dielectric passivation, metallization, and annealing steps to complete the SBD structure. First, the wafer surface was encapsulated with a conformal $SiO_2$ passivation layer deposited by plasma enhanced CVD in order to protect the etched sidewalls during subsequent processing and to define the active junction area. Standard optical lithography and $SiO_2$ etching were then used to open windows in the passivation over the recessed $n^+$ regions surrounding the mesa, where ohmic contacts to the heavily doped layer were formed. A Ti/Au metal stack was deposited by electron beam evaporation and patterned using lift-off to define these cathode contacts, followed by rapid thermal annealing at 470 °C for 1 minute in nitrogen ambient to improve contact resistivity. A second photolithography step was employed to open a circular window in the passivation directly above the central drift region at the mesa top, after which Ni/Au was deposited and lifted off to form the Schottky anode electrode. The final layout as shown in Figure 1(f) therefore consisted of a Ni/Au anode contacting the drift layer on top of the etched mesa, flanked by Ti/Au cathodes that accessed the underlying $n^+$ β-$Ga_2O_3$ layer on the planar regions.

Neutron irradiation of the completed SBD was performed at the University of Massachusetts Lowell Radiation Laboratory using the Fast Neutron Irradiator (FNI). The FNI provides a fast to thermal neutron flux ratio greater than 4000:1 and, at a reactor power of 1 MW, supplies a 1 MeV equivalent neutron flux of $8\times10^{10}$ n $cm^{-2}$ $s^{-1}$ with less than ±20 percent spatial variation across the sample plane. Irradiations were conducted at room temperature under unbiased condition and were timed to achieve a total fluence of $1\times10^{15}$ n $cm^{-2}$, corresponding to the calibrated 1 MeV equivalent flux of the FNI. After irradiation, the devices were transferred to a temperature-controlled probe station equipped with a resistive heating chuck and operated under air ambient. A single temperature cycle was performed in the sequence 25, 100, 200, and 250 °C, followed by cooling to 25 °C. The chuck temperature was increased at approximately 15°C/min, and the devices were held at each temperature for 5 min to reach thermal equilibrium before electrical characterization. The devices remained unbiased during heating and temperature stabilization and were electrically biased only during the measurements. At each temperature, current-voltage (J-V) measurements under both forward- and reverse-bias conditions were performed in the same sequence using a Keithley 4200A-SCS semiconductor parameter analyzer. Room-temperature reverse-bias measurements were extended to higher voltages using the same system to determine the breakdown voltage of devices with a 70 μm-diameter Schottky anode. Capacitance-voltage (C-V) measurements were performed at room temperature before irradiation, immediately after irradiation, and following completion of the electro-thermal cycle, using a frequency of 1 MHz and a DC-bias range from -20 to 0 V. Following the measurement at 250 °C, the devices were allowed to cool naturally to room temperature. Post-cycle room-temperature measurements were collected after the chuck stabilized at 25 °C.

## III. RESULTS AND DISCUSSION

Figure 2(a) presents the room temperature J-V characteristics of the SBDs before neutron irradiation, immediately after exposure to a fluence of $1\times10^{15}$ n $cm^{-2}$, and following an electro thermal annealing cycle. Prior to irradiation, the devices exhibit strong rectification, characterized by a turn-on voltage of 1.20 V (defined at a current density of 1 A/cm²), an ideality factor (η) of 1.32, and a minimum differential specific on-resistance ($R_{on,sp}$) of 8.43 mΩ cm², consistent with favorable forward transport in the LPCVD-grown structure and the plasma-free Ga-assisted mesa-etching process. Neutron irradiation leads to a degradation in forward conduction due to displacement damage within the β-$Ga_2O_3$ lattice. The turn on voltage increases to 2.40 V, the forward current is suppressed, and $R_{on,sp}$ increases to 8.85 mΩ cm², accompanied by a slight increase in the η from 1.32 to 1.36. Although the $R_{\mathrm{on,sp}}$ remains only slightly higher after irradiation, the forward J-V characteristics exhibit a pronounced degradation over the low- and intermediate-bias regions, as evidenced by the increased turn-on voltage and reduced current density at a given applied voltage. At higher forward bias, however, the irradiated and pre-irradiation J-V characteristics progressively converge, resulting in comparable $R_{\mathrm{on,sp}}$ values despite the overall degradation in forward conduction. The Schottky barrier height (SBH) extracted from forward J-V analysis was determined using the standard thermionic emission model [55],

$$J = J_s\left[exp\left(\frac{qV}{\eta k_0 T}\right) - 1\right] \qquad (1)$$

$$J_s = A^* T^2\ exp\left(-\frac{q\varphi_B^{JV}}{k_0 T}\right) \qquad (2)$$

$$A^* = \frac{4\pi q m_n^* k_0^2}{h^3} \qquad (3)$$

where q is the electric charge, $k_0$ is the Boltzmann constant, $J_s$ is the reverse saturation current density, $\varphi_B^{JV}$ is the apparent SBH obtained from the TE model [56], and A* is Richardson's constant, which is calculated to be 41.04 $Acm^{-2}K^{-2}$ [57]. Because neutron-induced traps can influence $\eta$ and thereby affect the apparent SBH obtained from the standard thermionic-emission relation, the Wagner correction was applied to account for these deviations and to provide a more accurate estimate of the true SBH [58].

$$\varphi_B = \left(\varphi_B^{JV} - \frac{\eta - 1}{\eta}\frac{kT}{q} ln\frac{N_C}{N_D}\right)\eta \qquad (4)$$

where $N_C$ is the effective conduction-band density of states. Using Eq. (4), $\Phi_B$ was determined to be 1.29 ± 0.02 eV before irradiation and 1.34 ± 0.02 eV after irradiation. This increase in $\Phi_B$ after irradiation mirrors the behavior reported in prior neutron-irradiation studies of β-$Ga_2O_3$ SBDs [49, 50] and could be attributed to the preferential suppression of low-barrier patches when radiation induced deep traps form in or near the depletion region [47, 59]. Following the electro-thermal cycle, the extracted $\Phi_B$ was 1.32 ± 0.02 eV, with minor changes observed in the minimum $R_{\mathrm{on,sp}}$ (8.85 to 8.72 mΩ·cm²) and $V_{\mathrm{on}}$ (2.40 to 2.30 V). Although the extracted $\Phi_B$ after electro-thermal cycling is slightly lower than the post-irradiation value, the difference is comparable to the experimental uncertainty and therefore does not constitute conclusive evidence of barrier-height recovery. Nevertheless, the overall trends in $V_{\mathrm{on}}$ and the minimum differential $R_{\mathrm{on,sp}}$ suggest limited recovery in the forward-transport characteristics following the electro-thermal cycle.

The physical origins of the irradiation-induced changes in $V_{\text{on}}$ and $R_{\text{on,sp}}$ can be attributed to the combined effects of donor compensation, carrier trapping, and modification of the effective Schottky barrier. Neutron-induced displacement damage introduces acceptor-like deep levels that compensate shallow donors, thereby reducing the free-electron concentration and increasing the drift-layer resistivity. Radiation-induced defects may also contribute to changes in the effective electron mobility through increased carrier scattering. In parallel, the increase in the $\Phi_B$ suppresses thermionic emission, while the increased drift and series resistances produce a larger voltage drop under forward conduction. These combined effects suppress the forward current over the low- and intermediate-bias regions and increase the voltage required to reach the reference current density used to define $V_{\text{on}}$. Because the extracted $\Phi_B$ increases by only 0.05 eV, the barrier-height change alone is unlikely to account for the full increase in $V_{\text{on}}$ from 1.20 to 2.40 V, indicating substantial contributions from donor compensation, increased series resistance, and defect-mediated carrier trapping. The comparatively small changes following the electro-thermal cycle suggest that the dominant compensating defects remain thermally stable under the conditions investigated.

The C-V characteristics shown in Figures 2(b) and (c) provide further insight into the electrical consequences of neutron irradiation. Before irradiation, the $1/C^2$-V characteristic exhibits the expected linear behavior associated with an approximately uniform net donor concentration in the drift layer. Immediately after irradiation, the capacitance decreases substantially. Following electro-thermal annealing, no significant change in the capacitance profile is observed within the measurement uncertainty, and the capacitance remains well below the pre-irradiation level. The built in potential $V_{bi}$ was extracted using the linear region of the $1/C^2$-V curve according to [55, 60]

$$\frac{A^2}{C^2} = \left[\frac{2}{q\varepsilon_r\varepsilon_0(N_D^+ - N_A^-)}\right]\left(V_{bi} - \frac{kT}{q} - V\right) \quad (5),$$

yielding $V_{bi}$ values of 1.16 V before irradiation, 1.18 V after irradiation, and 1.18 V following electro thermal annealing. The net donor concentration was determined using

$$N_D^+ - N_A^- = \frac{2}{q\varepsilon_r\varepsilon_0 A^2 (\frac{d\frac{1}{C^2}}{dV})} \quad (6)$$

Before irradiation, the donor concentration in the drift layer was measured to be $2.07\times10^{17}$ cm$^{-3}$. After neutron exposure, this value decreases to $1.12\times10^{17}$ cm$^{-3}$, representing a reduction of nearly 46 percent. This corresponds to a carrier-removal rate of ~95 cm$^{-1}$, consistent with strong donor compensation under fast-neutron displacement damage in β-$Ga_2O_3$ [49-51]. Following electro-thermal annealing, the extracted donor concentration remained essentially unchanged. Any apparent variation is within the experimental uncertainty of the C-V measurements and therefore does not constitute conclusive evidence of carrier-concentration recovery.

The extent of carrier compensation reported in the literature varies considerably. Our recent study on homoepitaxial β-$Ga_2O_3$ SBDs also demonstrated substantial donor compensation under the same fast-neutron irradiation conditions [61]. These observations suggest that the carrier-removal response may be influenced by multiple factors, including the epitaxial growth platform, initial donor concentration, epitaxial growth process, pre-existing defect states, dislocation density, residual strain, and interface-related defects.

Although the present electrical measurements do not directly identify the responsible defects, previously reported neutron-induced deep levels near $E_C$ - 0.78 eV, $E_C$ - 1.22 eV, and $E_C$ - 2.00 eV, some of which have been associated with vacancy-related defects and defect complexes, are plausible contributors to the observed donor compensation [47]. Separately, DLTS/DLOS studies of as-grown homoepitaxial β-$Ga_2O_3$ film grown by LPCVD have reported deep levels at around $E_C - 0.8$ eV, $E_C - 2.0$ eV, $E_C - 3.6$ eV, and $E_C - 4.4$ eV [62]. However, whether these defects are modified or newly generated by neutron irradiation in the present Si-doped heteroepitaxial LPCVD material remains to be determined through deep-level defect characterization.

Finally, the SBH extracted from C-V analysis follows trends similar to those from J-V extraction [56]. The barrier height was determined using

$$\varphi_B = V_{bi} + \frac{KT}{q} ln\left[\frac{N_C}{N_D^+ - N_A^-}\right] \quad (7),$$

The extracted $\Phi_B$ increases from 1.25 eV before irradiation to 1.28 eV after irradiation and remains at 1.28 eV following the electro-thermal cycle. The similar trends obtained from the J-V and C-V analyses are consistent with neutron-induced changes in the carrier concentration and defect distribution within the drift region.

Temperature-dependent forward characteristics before and after neutron irradiation are shown in Figure 3(a). In the pre-irradiation condition, the forward current density increases steadily with temperature in the low to moderate forward-bias regime, consistent with thermionic-emission-dominated transport in $\beta$-$Ga_2O_3$ [26, 63], where increased thermal energy enhances electron injection over the Schottky barrier. After irradiation, a reduction in forward current is observed at all temperatures, reflecting neutron-induced carrier removal and the formation of deep compensating centers within the drift region. Although the overall temperature trend remains similar to the pristine condition, the irradiated diodes show consistently lower current density at any given temperature, indicating that neutron-induced displacement damage suppresses electron injection [50] and increases the effective series resistance.

In addition to the forward-bias behavior, the reverse-bias characteristics in Figure 3(a)

also show important trends with temperature and irradiation. For the pre-irradiation devices, the reverse current increases steadily as temperature rises from 25 to 250 °C, consistent with thermally activated transport mechanisms in β-$Ga_2O_3$, including thermionic field emission and trap-assisted leakage through the depletion region. As the temperature increases, a greater fraction of electrons acquires sufficient thermal energy to overcome or tunnel through the barrier [26, 64], resulting in higher leakage current. After neutron irradiation, the reverse current remains consistently lower than in the pristine devices across the entire temperature range. This suppression of leakage is characteristic of donor compensation, since the reduction in free carrier concentration enlarges the depletion width and alters the electric-field distribution within the drift layer. The modest increase in SBH extracted from both J-V and C-V analysis further confirms this behavior by raising the effective energy barrier for electron emission.

The temperature-dependent $R_{on,sp}$ is plotted in Figure 3(b). Before irradiation, $R_{on,sp}$ increases from 8.43 mΩ·cm² at 25 °C to 10.13 mΩ·cm² at 250 °C. This behavior is characteristic of β-$Ga_2O_3$ drift conduction, where the dominant mobility-limiting mechanism is polar optical phonon scattering, which strengthens with temperature and lowers electron mobility [65]. After neutron irradiation, $R_{on,sp}$ exhibits a uniform upward shift across the entire temperature range. At 25 °C, $R_{on,sp}$ increases to 8.85 mΩ·cm², and continues to rise to 9.28 mΩ·cm² at 100 °C, 10.12 mΩ·cm² at 200 °C, and 10.85 mΩ·cm² at 250 °C. This increase of $R_{on,sp}$ at every temperature is consistent with the reduction of donor concentration extracted from C-V profiling. Figure 3(c) shows the extracted $R_{on,sp}$ values at each temperature, indicating the consistent separation between the pre- and post-irradiation curves. The monotonic increase of $R_{on,sp}$ with temperature in both cases indicates that thermionic emission remains the dominant transport mechanism.

Figure 4(a) presents the temperature dependence of the η and $\Phi_B$ before and after neutron irradiation. In the pre-irradiation condition, η increases gradually from 1.32 at 25 °C to 1.42 at 250 °C. This rise with temperature is consistent with thermionic-emission-dominated transport in β-$Ga_2O_3$, where increased thermal energy activates additional transport channels such as barrier inhomogeneity and limited recombination-assisted flow [66, 67]. Following neutron irradiation, η is elevated across the entire temperature range, increasing from 1.36 at 25 °C to 1.48 at 250 °C. The temperature dependence of the $\Phi_B$ , also shown in Figure 4(a), exhibits characteristic trends for both pre- and post-irradiation devices. Before irradiation, $\Phi_B$ decreases slightly from 1.29 eV at 25 °C to 1.25 eV at 250 °C, consistent with the negative temperature coefficient associated with lateral barrier inhomogeneity in β-$Ga_2O_3$ [63, 68]. Following irradiation, the extracted $\Phi_B$ remains higher than the corresponding pre-irradiation value across the measured temperature range, while decreasing from 1.34 eV at 25 °C to 1.28 eV at 250 °C. This similar temperature dependence suggests that barrier inhomogeneity may remain relevant after irradiation, although the defect and carrier distributions are modified.

The reverse leakage current in Figure 4(b) shows the strong thermal activation behavior expected for β-$Ga_2O_3$ SBDs [69, 70]. In the pristine device, the leakage current rises from $4.94\times10^{-6}$ A/cm² at 25 °C to $1.71\times10^{-4}$ A/cm² at 250 °C, an increase by over two orders of magnitude. After neutron irradiation, however, the leakage is significantly reduced at every temperature. At 25 °C, the leakage falls to $5.05\times10^{-7}$ A/cm², nearly an order of magnitude lower than the pristine case. At 100 °C and 200 °C, the leakage currents are reduced from $1.31\times10^{-5}$ A/cm² to $3.24\times10^{-6}$ A/cm² and from $2.91\times10^{-5}$ A/cm² to $5.48\times10^{-6}$ A/cm², respectively. Even at 250 °C, the irradiated device shows a leakage current of $2.10\times10^{-5}$ A/cm², which is still an order of magnitude lower than the $1.71\times10^{-4}$ A/cm² measured before irradiation. This consistent suppression of leakage current could arise from the donor-compensating nature of neutron-induced defects, which reduce the free electron concentration in the drift region [50]. The concurrently increased $\Phi_B$ confirms this behavior by further reducing the thermionic emission probability.

The rectification ratio $I_{on}/I_{off}$ reflects the combined effects of forward and reverse transport. Before irradiation, $I_{on}/I_{off}$ decreases from $1.49\times10^7$ at 25 °C to $1.15\times10^6$ at 250 °C, predominantly driven by the sharp increase in leakage current at elevated temperatures. After irradiation, the rectification ratio is reduced to $1.08\times10^7$ at 25 °C because the forward current is reduced more strongly than the reverse current. Nevertheless, the temperature degradation of $I_{on}/I_{off}$ after irradiation is more gradual than in the pristine device. At 100 °C, $I_{on}/I_{off}$ shifts from $4.34\times10^6$ (pre) to $4.45\times10^6$ (post), showing that the irradiated device maintains similar rectification at this temperature despite its reduced carrier concentration. At 200 °C, $I_{on}/I_{off}$ decreases from $3.69\times10^6$ to $2.55\times10^6$, and at 250 °C, from $1.15\times10^6$ to $5.28\times10^5$. This slower degradation in rectification ratio with temperature arises because the irradiated device maintains substantially lower leakage at all temperatures, offsetting the reduced forward current and resulting in a narrower spread between room-temperature and high-temperature rectification performance.

Figure 5(a) shows the reverse J-V characteristics before and after neutron irradiation. In the pre-irradiation condition, the devices exhibit a breakdown voltage of 101 V for the moderately doped drift layer ($2.07\times10^{17}$ cm$^{-3}$). After exposure to a neutron fluence of $1\times10^{15}$ n cm$^{-2}$, the breakdown voltage increases to 135 V. This enhancement can be correlated with the neutron-induced donor compensation observed in the C-V analysis, where the net donor density decreases from $2.07\times10^{17}$ cm$^{-3}$ to $1.12\times10^{17}$ cm$^{-3}$ [45, 50, 61]. The reduced donor concentration leads to an expansion of the depletion width under reverse bias, causing the electric field to be distributed over a larger spatial region and allowing the device to sustain a higher reverse voltage before breakdown occurs [71]. Using the one-dimensional electrostatic estimate for the maximum electric field [55], the pre-irradiation breakdown voltage of 101 V at a donor concentration of $2.07\times10^{17}$ cm$^{-3}$ corresponds to a peak field of 2.75 MV·cm$^{-1}$. After irradiation, the combination of reduced donor density ($1.12\times10^{17}$ cm$^{-3}$) and slightly increased $V_{BR}$ of 135 V yields a lower estimated peak field of 2.34 MV·cm$^{-1}$. No measurable change in $V_{\mathrm{BR}}$ was observed following the electro-thermal cycle, consistent with the negligible change in the net donor concentration.

Silvaco TCAD simulations provide further insight into the electric-field distribution and breakdown mechanism following neutron irradiation. The simulations were performed using the experimental quasi-vertical β-$Ga_2O_3$ SBD geometry. The simulated structure comprised a 1.05 μm unintentionally doped buffer layer, a 2.10 μm $n^+$ contact layer, a 3.15 μm drift layer, and a 3.6 μm mesa recess, with Ni/Au Schottky and Ti/Au ohmic contacts. The lateral dimensions were selected to match the fabricated device shown in Fig. 1. The pre- and post-irradiation drift-layer net donor concentrations were set to $2.07 \times 10^{17}$ and $1.12 \times 10^{17}$ cm$^{-3}$, respectively, based directly on the C-V measurements. Neutron-induced compensation was therefore represented by the measured reduction in net donor concentration rather than by explicitly introducing defect levels. Because the β-$Ga_2O_3$/$SiO_2$ interface fixed charge was not independently measured, it was not included in the simulations. The same interface condition was used for the pre- and post-irradiation structures, allowing the effect of the C-V-extracted change in net donor concentration on the electric-field distribution to be evaluated independently of any assumed interface-charge variation. The simulations employed concentration-dependent mobility, field-dependent mobility, Fermi-Dirac carrier statistics, and Shockley-Read-Hall recombination models using the standard TCAD model library. The same physical models were used for both the pre- and post-irradiation simulations. The β-$Ga_2O_3$ parameters included a relative permittivity of 10.0, electron and hole mobilities of 150 and 1 $cm^2 \cdot V^{-1} \cdot s^{-1}$, respectively, and effective conduction- and valence-band densities of states of $3.7 \times 10^{18}$ and $2.5 \times 10^{18}$ $cm^{-3}$, respectively [72, 73]. Figures 5(b-d) show the simulated electric-field distributions at the experimentally measured breakdown voltages of 101 V (pre-irradiation) and 135 V (post-irradiation). In both cases, the highest electric field occurs at the Schottky anode edge (points B and B′), consistent with a corner-limited breakdown mechanism. After irradiation, the reduced donor concentration widens the depletion region and results in a more spatially distributed electric-field profile near the anode edge. Quantitatively, the simulated peak electric field at the Schottky edge decreases from 2.89 $MV \cdot cm^{-1}$ before irradiation to 2.48 $MV \cdot cm^{-1}$ after irradiation, while the drift-region field at the device center (point D) decreases from 2.58 to 2.13 $MV \cdot cm^{-1}$. These trends are consistent with the one-dimensional electrostatic estimates and indicate that neutron-induced donor compensation reduces field crowding at the anode edge and lowers the maximum electric field within the drift region.

## IV. CONCLUSION

In summary, we have investigated the effects of fast-neutron irradiation on heteroepitaxial Ni/β-$Ga_2O_3$ SBDs on c-plane sapphire substrates. Neutron exposure at a high fluence of $1 \times 10^{15}$ n $cm^{-2}$ leads to a reduction in forward current, an increase in turn-on voltage, and suppressed reverse leakage, consistent with neutron-induced donor compensation and defect-mediated changes within the drift region. The net donor concentration decreases by ~46 percent, corresponding to a carrier-removal rate of ~95 $cm^{-1}$ after irradiation, while the barrier height increases. Temperature-dependent measurements show that thermionic emission remains the dominant transport mechanism after irradiation. A slight increase in breakdown voltage is observed after irradiation due to the reduced net carrier concentration, with the overall reverse characteristics remaining stable. Overall, these results provide insight into neutron-induced donor compensation in heteroepitaxial β-$Ga_2O_3$ and contribute to the understanding of radiation effects in $Ga_2O_3$ devices fabricated on non-native substrates.

**Acknowledgement**

The authors acknowledge the funding support from the Air Force Research Laboratory Regional Network- Mid-Atlantic and the National Science Foundation (NSF) under award numbers ECCS-2532898 and ECCS-2501623.

**Data Availability**

The data that support the findings of this study are available from the corresponding author upon reasonable request.

**Conflict of Interest**

The authors have no conflicts to disclose.

**Figure 1**

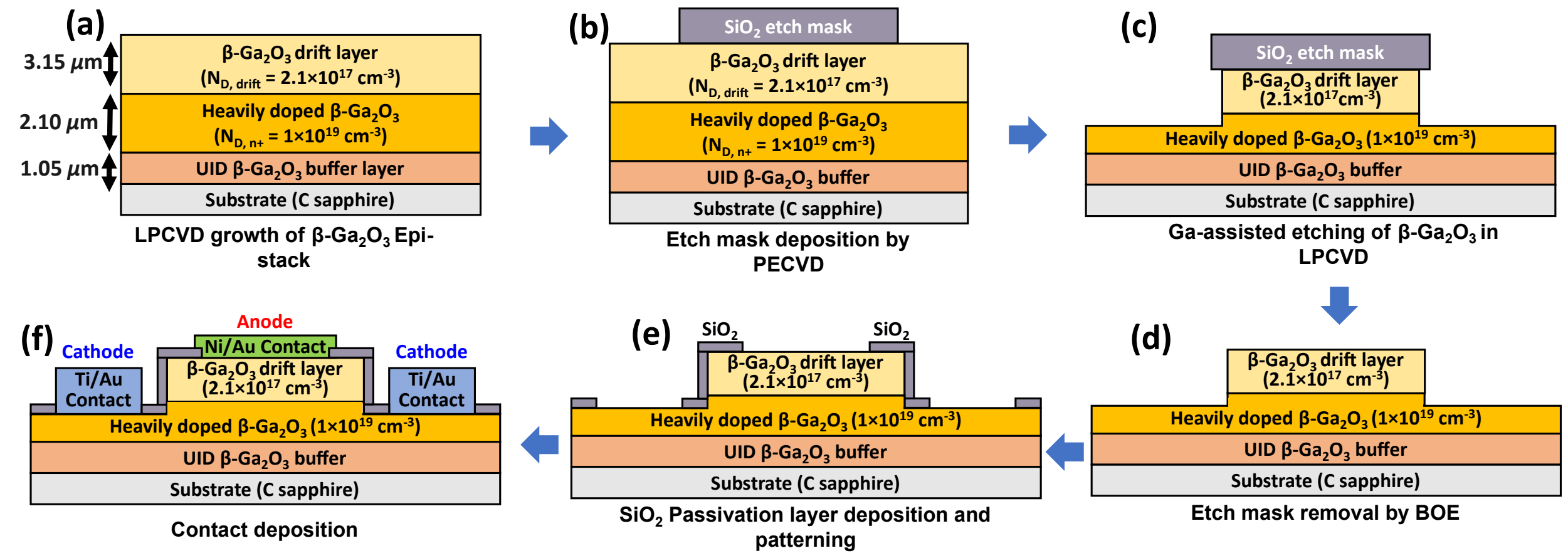

**Fig. 1.** (a)-(f) Process flow for fabricating quasi-vertical β-$Ga_2O_3$ SBDs on c-plane sapphire using an all-LPCVD growth and etching platform.

Figure 2

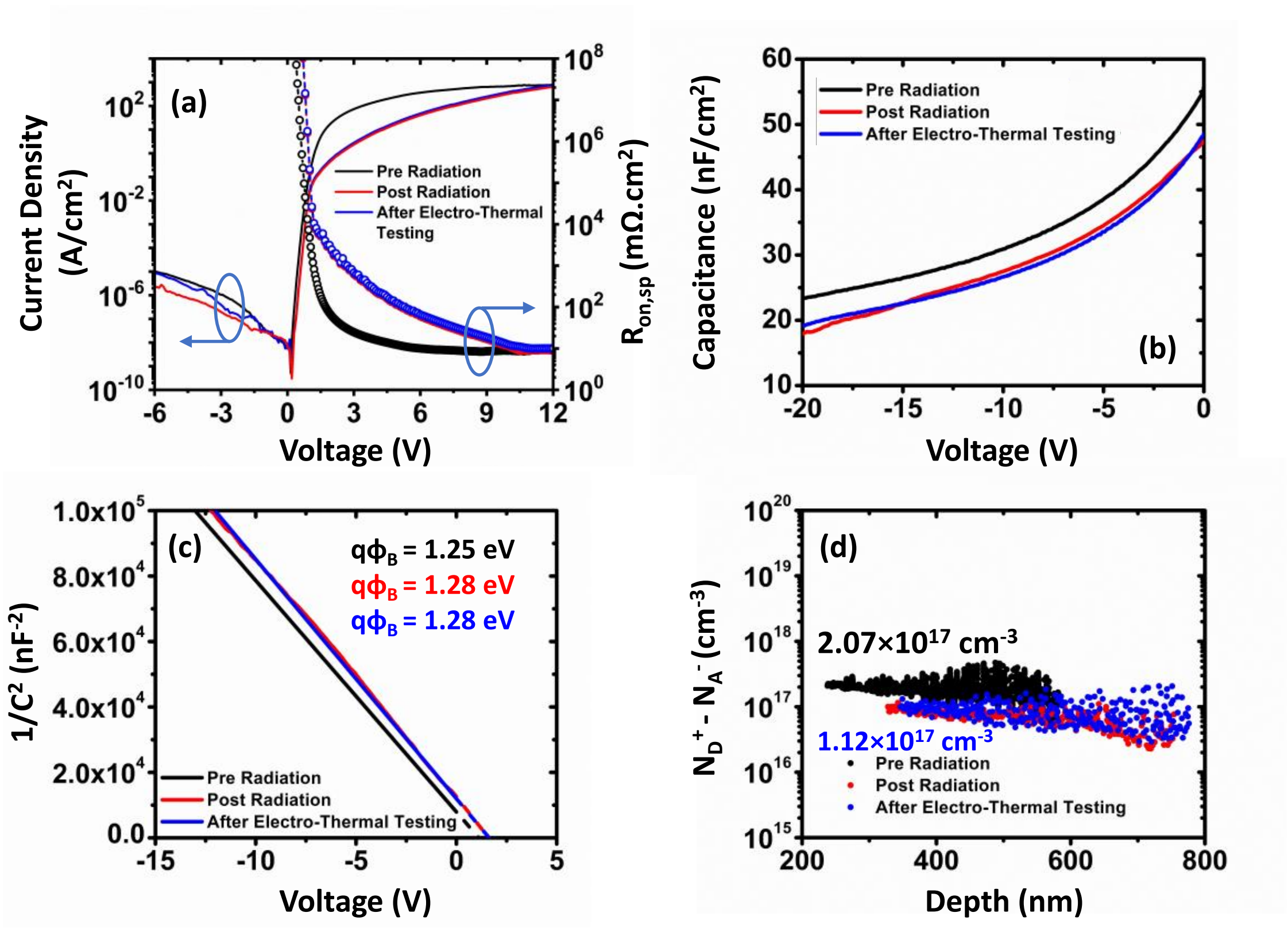

**Fig. 2.** Room-temperature electrical characteristics of LPCVD-grown β-$Ga_2O_3$ SBDs before and after fast-neutron irradiation (1 MeV equivalent fluence of $1\times10^{15}$ n cm$^{-2}$) and following an electro-thermal annealing cycle. (a) Forward J-V characteristics. (b) C-V characteristics. (c) Corresponding $1/C^2$-V plots used to extract the built-in potential and Schottky barrier height. (d) Net donor concentration profiles extracted from the C-V measurements.

## Figure 3

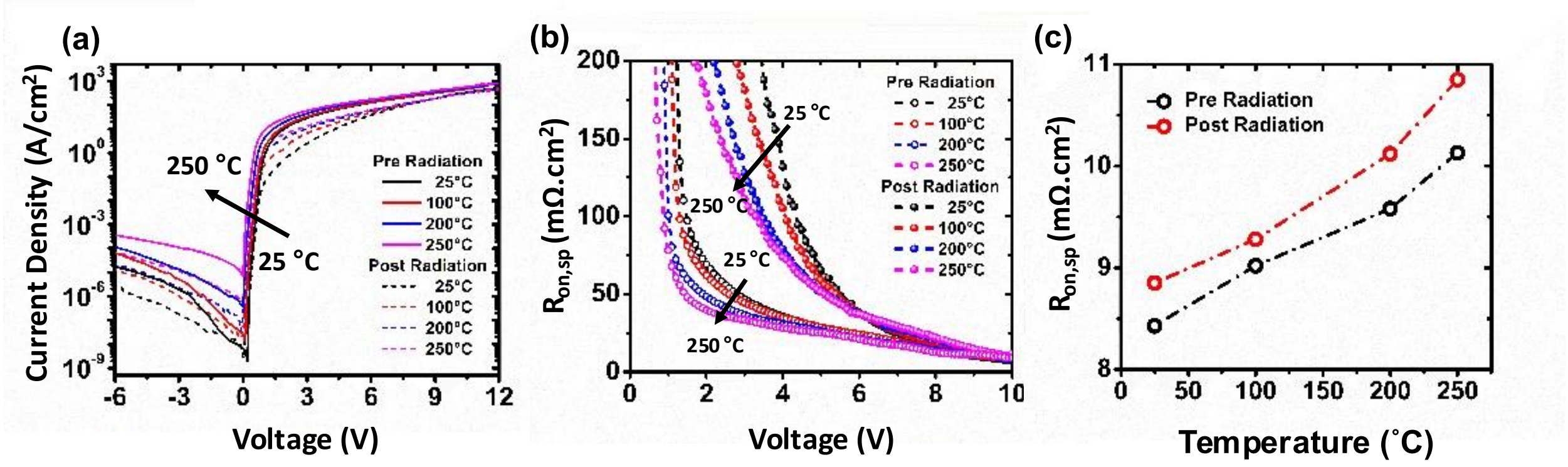

**Fig. 3.** Temperature-dependent electrical characteristics of β-$Ga_2O_3$ SBDs before and after fast-neutron irradiation. (a) Forward J-V characteristics measured from 25 to 250 °C. (b) Temperature-dependent specific on-resistance ($R_{on,sp}$). (c) Comparison of minimum $R_{on,sp}$ before and after irradiation.

**Figure 4**

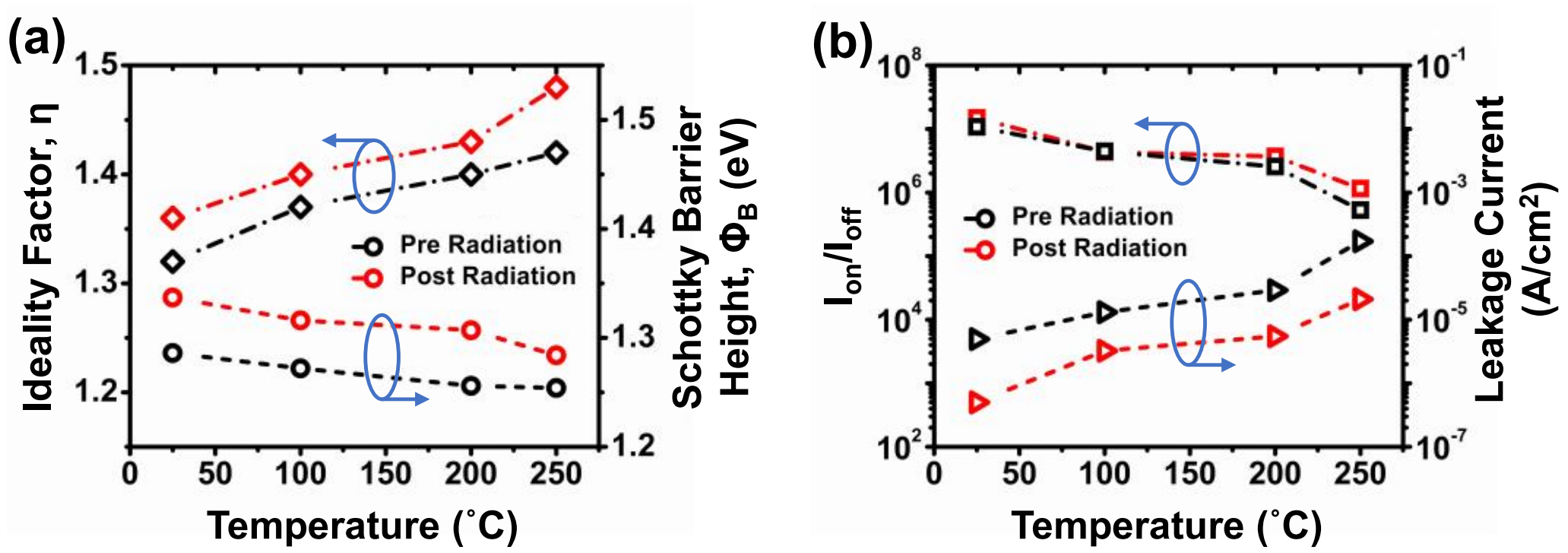

**Fig. 4.** Impact of neutron irradiation on temperature-dependent ideality factor, Schottky barrier height, reverse leakage, and rectification ratio. (a) Temperature dependence of the ideality factor η and Schottky barrier height $\Phi_B$ extracted from forward J-V characteristics before and after irradiation. (b) Reverse leakage current density and rectification ratio as a function of temperature for pristine and irradiated devices.

**Figure 5**

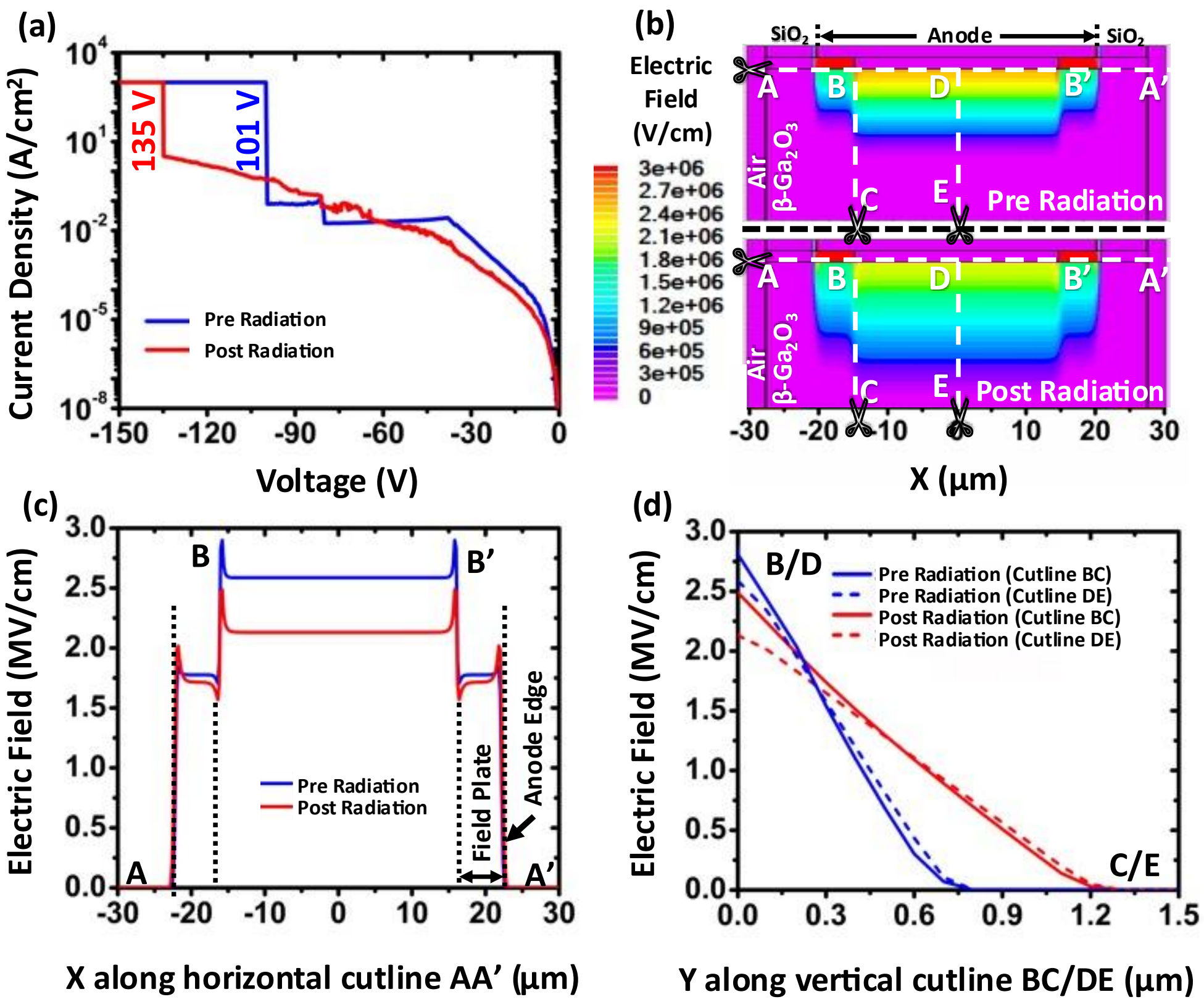

**Fig. 5.** (a) Reverse J-V characteristics of the diodes before and after neutron irradiation. (b) Simulated electric field distribution under reverse bias at the respective experimental breakdown voltages. (c) Horizontal electric-field profiles along AA′ showing field maxima associated with the Schottky metal edge. (d) Vertical field profiles along BC and DE demonstrating the wider depletion region and redistributed electric field region in the compensated post-radiation device.